\documentclass[
twocolumn,
showpacs,amssymb,aps,prl]{revtex4}

\usepackage{graphicx}
\usepackage{bm}

\begin{document}
\title{
Influence of a Parallel Magnetic Field on Microwave Photoconductivity\\
 in a High-Mobility 2D Electron System 
}
\author{C. L. Yang}
\author{R. R. Du}
\affiliation{Department of Physics, University of Utah, Salt Lake City, Utah 84112}
\affiliation{Department of Physics and Astronomy, Rice University, Houston, Texas 77005}
\author{L. N. Pfeiffer}
\author{K. W. West}
\affiliation{Bell Laboratories, Lucent Technologies, Murray Hill, New Jersey 07974}
\begin{abstract}
We have studied experimentally the influence of a parallel magnetic field ($B_{//}$) on 
microwave-induced resistance oscillations (MIRO) and zero-resistance states (ZRS) 
previously discovered in a high-mobility 2D electron system. We have observed a strong 
suppression of MIRO/ZRS by a modest $B_{//}\sim 0.5$ T. In Hall bar samples, magnetoplasmon
resonance (MPR) has also been observed concurrently with the MIRO/ZRS. In contrast to 
the suppression of MIRO/ZRS, the MPR peak is found to be enhanced by $B_{//}$. These findings have 
not been addressed by current models proposed to explain the microwave-induced effects. 

\end{abstract}
\maketitle

Microwave induced resistance oscillations (MIRO)~\cite{zudovprb} and zero-resistance states 
(ZRS)~\cite{mani, zudovzrs}, observed in a high mobility two-dimensional electron system (2DES) 
in GaAs/AlGaAs heterostructures, have been the subject of intense current interest in the condensed matter 
community~\cite{du_zrs}. It has been theoretically shown~\cite{andreev} that the ZRS can be the consequence 
of an instability due to  absolute negative conductivity (ANC) occurring at the oscillation minima 
when the MIRO are sufficiently strong. The instability leads to the formation of electric current
domains with zero voltage, resulting in a measured zero resistance. In this context, a large number of
microscopic mechanisms have been proposed to account for the MIRO and the subsequent ANC, based on 
microwave-assisted electron scattering~\cite{durst} or microwave-induced non-equilibrium electron 
distribution~\cite{dorozhkin, dmitriev}  and the oscillatory density of states due to Landau quantization. 
Other interesting mechanisms, such as orbit self-organization~\cite{phillips}, or quantum interference 
effect~\cite{DHLee}, have been proposed. 

Recently, the importance of electrodynamic effects has been emphasized~\cite{Mikhailov2004, kukushkin, studenikin} 
for the microwave response of a high mobility 2DES. Specifically, the plasmon modes are expected to play a role on the 
MIRO and ZRS in real samples with finite size. It is well known that, for a 2DES with lateral width $w$, the (bulk) 
magnetoplasmon resonance (MPR) occurs at~\cite{review_MP, MPR_vasiliadou}
\begin{equation}
\omega^2=\sqrt{\omega_c^2+\omega_0^2},
\end{equation}
where $\omega=2\pi f $ is the MW frequency, 
$\omega_c=eB/m^*$ is the cyclotron frequency,
and
$\omega_0 =\sqrt{n_ee^2q/2\epsilon\epsilon_0m^*}$
is the 2D plasma frequency, with an effective dielectric constant
$\epsilon=(1+\epsilon_{GaAs})=6.9$ for GaAs, and a plasma wave vector
$q=\pi/w$. The MPR has been observed, together with MIRO, in microwave photoconductivity
of lower-mobility 2DES ($\mu=3\times 10^6$ cm$^2$/Vs)~\cite{zudovprb}. However, 
so far it has not been reported in ultra-clean samples where ZRS are observed. This has become a issue relevant 
to the origin of the MIRO and ZRS~\cite{Mikhailov2004}.      

In this paper, we report on the influence of a parallel magnetic field on the microwave photoconductivity of 
ultra-clean 2DES. The MIRO and ZRS, as well as the magneto-plasma resonance, are experimentally studied 
under a fixed parallel magnetic field ($B_{//}$) and sweeping perpendicular field ($B_z$),
with $B_z$ and $B_{//}$ independently provided by a two-axes magnet.
Our main finding is that the oscillatory photoconductivity is strongly suppressed by a 
relatively small $B_{//}$, of the order of 0.5 T.
And we found that the MPR can occur \emph{concurrently} with the MIRO and ZRS, while its strength
is \emph{enhanced} by $B_{//}$ in contrast to the suppression of MIRO and ZRS.
In addition, employing $B_{//}$ we were also able to uncover a new, resistively detected resonance 
which occurs near twice the cyclotron frequency. 
These new experimental findings in a parallel magnetic field have not 
been addressed by current theoretical work on the photo-response of  high mobility 2DES,
and could help to differentiate between competing theories.

Ultra-clean 2DES used in these experiments were provided by 
Al$_{0.24}$Ga$_{0.76}$As/GaAs/Al$_{0.24}$Ga$_{0.76}$As square quantum wells (QW); 
similar data have been obtained from single-interface GaAs/Al$_{0.3}$Ga$_{0.7}$As heterostructures. 
Data reported in this paper are from two specimen, sample $QW30nm$ ($QW25nm$), which is a 30-nm (25-nm) 
width QW, having an electron density of $n_e = 3.45\times10^{11}$ ($3.30\times 10^{11}$) cm$^{-2}$ and a mobility of
$\mu = 1.8\times10^7$ ($1.9\times10^7$) cm$^2$/Vs at 0.35 K. These parameters were obtained after a 
brief illumination from a red light-emitting diode at 2 K. Both specimen were lithographically defined and 
wet-etched: the sample $QW30nm$ is a 200 $\mu$m x 400 $\mu$m rectangle and the sample QW25nm is 
a 100 $\mu$m x 2 mm Hall bar. The measurements were performed in a top-loading He-3 cryostat with a 
base temperature of 0.35 K; microwaves were generated by Gunn diodes and guided down to the 
sample (Faraday configuration) via a WR-28 waveguide. The long-side of the samples was placed in parallel with 
the long-side of the  waveguide cross section (the direction is denoted by $x$)
Four terminal, low frequency (23 Hz) lock-in technique with a excitation current $I = 1\ \mu$A was employed. 
A two-axes Nb superconducting solenoid system was employed to provide a perpendicular field, $B_z$ (up to 3 T) 
and a parallel field, $B_{//}$ (up to 2 T). 
\begin{figure}
\includegraphics{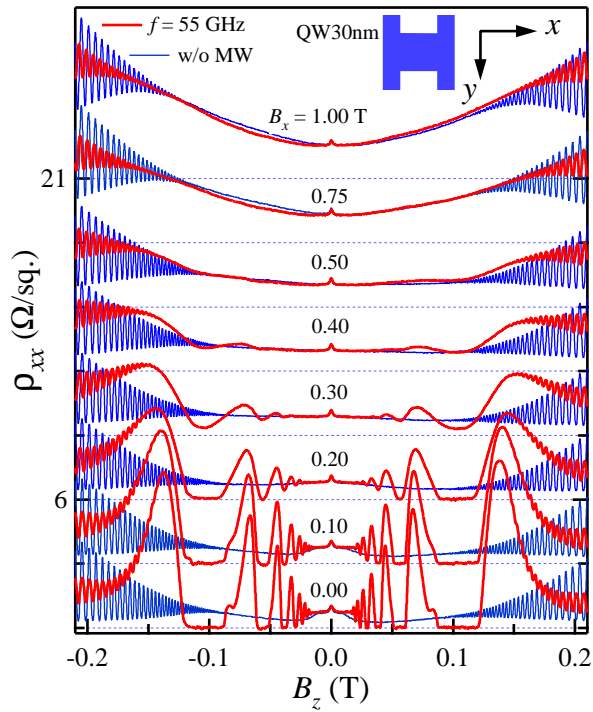}
\caption{Microwave-induced resistance oscillations (MIROs) and the zero-resistance states (ZRS)
observed on a ultra-high-mobility 2DES (sample QW30nm, see text) 
subjected to parallel magnetic field $B_x$ (for clarity, 
traces are vertically shifted by steps of 3 $\Omega$/sq.).  
For comparison, corresponding traces without microwave irradiation (thin lines) are also shown. 
}
\end{figure}

An overview of MIRO and ZRS (thick lines) in a parallel magnetic field $B_{x}$ for sample $QW30nm$ is shown in FIG.1.  
The traces were recorded at a temperature $T\approx0.95$ K while the smple was continuously irradiated by a microwave 
of frequency $f=55$ GHz and power (on the sample surface) $P\approx100\ \mu$W. For comparison, corresponding 
magnetoresistivity (MR) without MW (at $T\approx0.35$ K), is also shown (thin lines). 
At $B_{//}=0$, the ``dark'' traces show negative MR around $B_{z}=0$  
which consist of a bump ($|B_{z}|\lesssim 0.025$ T) and a sharp peak ($|B_{z}|\lesssim 0.003$ T) on top of it.  
In general, with increasing $B_{//}$ the bump of negative MR disappears rapidly and only the sharp peak close to 
$B_{z}=0$ is left. For all $B_{//}$ applied (up to 2.0 T), sharp Shubnikov--de Haas (SdH) oscillations are observed 
while the onset moves to higher magnetic field with increasing $B_{//}$ (from $B_z\sim0.08$ T at $B_{//}=0$ to  
$B_z\sim0.12$ T at $B_{//}=1.0$ T). With MW irradiation, the most salient feature is that the MIRO diminish rapidly 
with increasing $B_{x}$: at $B_x = 0.2$ T, the amplitude has already decreased more than 50\% as compared to that 
at $B_{x} = 0$; and by $B_{x} = 0.75$ T, the oscillations completely disappear. Initially, for $B_{x} = 0$, up to four
ZRS are observed  at the mimima of the MIRO. Among them, the strongest first two ZRS (centered around 
$B_z = 0.10$ T and $B_z = 0.055$ T) persist up to $B_{x} = 0.2$ T, and then are quickly suppressed 
at $B_{x}= 0.3$ T.  Before the diminishing of the ZRS, shrinkage of their width
is noticed, for instance, at $B_{x} = 0.2$ T. 
We note that, with increasing $B_{x}$, the positions of the oscillation maxima shift towards higher $B_z$ 
while those of the ZRS and oscillation minima are roughly constant. 

By rotating the sample probe $90^{\circ}$, we were able to perform the same measurements 
with a parallel magnetic field $B_y$ which is perpendicular to the excitation current. The observations are 
qualitatively the same as those of FIG. 1. However,  the suppression of MIRO and ZRS
by $B_y$ is not as strong as that by $B_x$: The ZRS disapper at $B\lesssim 0.5$ T and the MIRO are complete 
suppressed at $B\lesssim1.5$ T. Despite of these quantitative difference, out main finding is that ZRS is strongly
suppressed by a modest parallel magnetic field $B_{//}$. 
\begin{figure}
\includegraphics{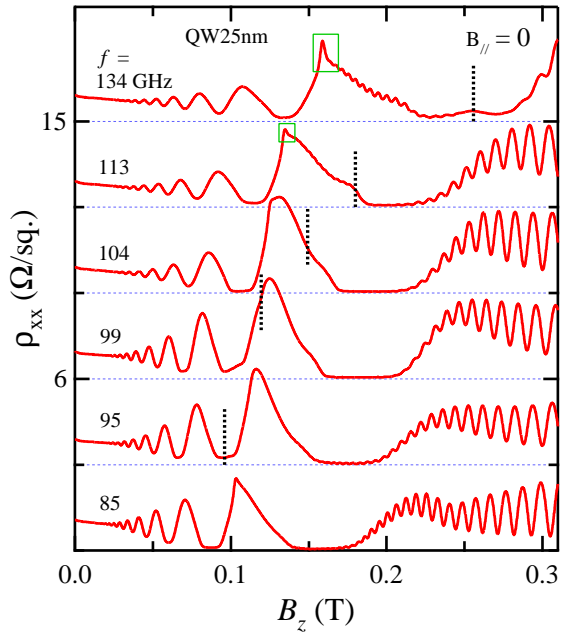}
\caption{\label{fig2}
Microwave response of sample QW25nm (described in text) recorded at $B_{//}=0$ for selected MW frequencies 
(traces are vertically shifted by steps of 3 $\Omega$/sq.). 
The dashed vertical lines indicate the positions of the magneto-plasma peak, as determined from traces 
taken at a $B_{//}$ where the magneto-plasma peak is prominent and easily resolved (see FIG. 3). 
The $f=134$ GHz and $f=113$ GHz traces show an extra peak (framed) on top of the second MIRO maximum,  
this extra peak generally appears for a wide range of MW frequencies and is better resolved when MIRO 
are suppressed by $B_{//}$ (cf. Fig. 3).
}
\end{figure}

In order to study the influence of $B_{//}$ on the resistively-detected MPR, we have performed similar 
experiments on the Hall bar sample of QW25nm. Not surprisingly, we have also observed a strong suppression of
MIRO/ZRS in this sample~\cite{foot_anisotropy}. However, the MPR peak is enhanced (rather than suppressed) by 
$B_{//}$. Typically, the magnetoplasmon peak is masked by the MIRO at $B_{//}=0$ but becomes dominant  
at elevated  $B_{//}$. As shown in Fig. 2, at $B_{//}=0$, with low  MW frequency the MPR
just slightly modifies the shape of the MIRO and can only be clearly identified when its position moves up 
into the wide first minimum of the MIRO, with increasing MW frequency. The enhancement of the MPR 
peak in an increasing $B_{//}$, concurrent with the suppression of the MIRO, is shown in the left panel of Fig. 3,  
for $f = 120$ GHz as an example.
Another important observation is that the peak position of the magneto-plasma mode is almost independent 
of $B_{//}$ (within the accuracy of  $1\%$) as indicated by the vertical line $A$ of the left panel of FIG. 3.  
This  means the effective electron mass ($m^*$) is nearly constant within the $B_{//}$ range studied.

The MP modes at various MW frequencies have also been explored,
example traces recorded at $B_y=1.50$ T for selected microwave frequencies are shown in the right panel of FIG. 3.
A set of magneto-plasma peak positions, plotted in the inset of FIG. 3 (right panel), were accurately determined by 
employing appropriate $B_{//}$ to simultaneously enhance the  magnetoplasmon mode and suppress the MIRO.
The fit to the dispersion relation Eq. (1) is excellent and it yields  $m^*=0.070\ m_e$, and 
a $f_0\equiv\omega_0/2\pi=87.0$ GHz which is 	reasonably close to the calculated value of 93 GHz. 
\begin{figure}
\includegraphics{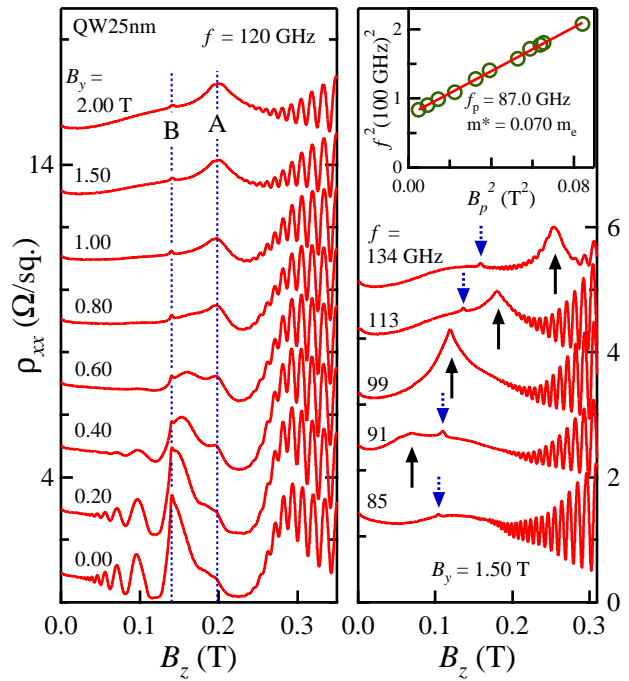}
\caption{\label{fig3}
Magneto-plasma resonance observed on sample QW25nm subjected to a parallel magnetic field. 
Left panel: Magneto-resistance traces taken at $f = 120$ GHz and diffevrent $B_y$s (vertically shifted 
with steps of 2 $\Omega$/sq.). In contrast to the diminishing of the MIRO oscillations, 
the Magneto-plasma resonance (vertical line $A$) becomes stronger with increasing $B_y$; in addition, another 
weak peak at $\omega/\omega_c\approx2$ (vertical line $B$) is also survived at increased $B_y$.  Right panel:  Magneto-resistance traces taken at $B_y=1.50$ T and different MW frequencies
(vertically shifted with steps of 1 $\Omega$/sq.). Magneto-plasma resonance peaks are indicated by up arrows,
and the weak peak at $\omega/\omega_c\approx2$ are indicated by down arrows. Insert: The plot 
of $f^2$ vs. $B_p^2$, which reveals the typical magneto-plasma dispersion relation. 
}
\end{figure}

In addition,  we observed a small sharp peak (marked by the vertical line $B$ in FIG. 3a) which overlaps with the 
second maximum of the MIRO; the $B_z$ position of this peak does not change with $B_{//}$ and its 
strength does not have strong dependence on $B_{//}$. Although the small peak is discernible for certain frequencies 
at $B_{//}=0$ (see FIG. 2, for example), it can be recognized as an extra peak only in elevated $B_{//}$. 
This new peak has been observed in a wide 
MW frequency range (50-150 GHz) and is not correlated with the magnetoplasmon mode (see the right 
panel of FIG. 4, the small peak appears at $f=85$ GHz where the magneto-plasma mode is absent).
Also the extra peak can not be identified as electron-spin resonance (ESR) since its position is not shifted
by $B_{//}$, and the ESR should occur at much higher magnetic field (at the order of 10 T) within the MW 
frequency range we used. Note this peak is distinct from MIRO by its behavior in $B_{//}$. Although the nature 
of this resistance peak is not clear, such a feature may signal a new resonance taking place near twice the cyclotron frequency, i.e., 
$\epsilon\equiv\omega/\omega_c\approx2$ with $\omega_c=eB_z/m^*$. 
However, the new resonance can not be simply regarded as a cyclotron harmonic because 
such a sharp feature is absent from the 1st MIRO peak.

Now we comment on possible mechanism leading to  the strong suppression of MIRO/ZRS by a $B_{//}$. 
In a strictly 2DES, a pure $B_{//}$ can only couple to the spins of the electrons through the Zeeman  
interaction. In a \emph{quai}-2DES with finite thickness, the $B_{//}$ can couple to the orbital motion of the electrons,
giving rise to a diamagnetic energy shift and a slightly increased effective mass in the direction perpendicular to  
$B_{//}$~\cite{Stern}; $B_{//}$ can also couple to the spins through spin-orbit interactions. These effects, however, 
hardly affect the Landau quantization under a perpendicular magnetic field $B_z$, at least within the parameter regime of our 
experiments; in particular, the Landau levels are still equally spaced and the oscillatory structure of the density of states is 
essentially intact. In this single-particle picture, it is not likely that the $B_{//}$ can significantly alter the photon-assisted 
scattering processes between Landau levels.  Therefore, the strong suppression of the MIRO/ZRS by $B_{//}$ may indicate a 
many-body or collective origin of the microwave-induced oscillations.

In summary, we have observed a strong suppression of the microwave-induced resistance 
oscillations and the subsequent zero-resistance states by a parallel magnetic field.
On the contrary, magnetoplasmon resonance peaks, observed concurrently with MIRO/ZRS, 
are robust in a parallel magnetic field. 

We acknowledge helpful discussions with A. H. Mac Donald, F. von Oppen, M. E. Raikh, and M. G. Vavilov.
C. L.Y. and R. R. D. were supported by NSF DMR-0408671.

\bibliographystyle{unsrt}

\begin{thebibliography}{10}
\bibitem{foot_anisotropy}
Contrary to sample QW30nm, the effect of $B_{//}$ on MIRO/ZRS in sample QW25nm does 
not show significant anisotropy with respect to the current direction, $x$; 
data shown in FIG. 3 are for $B_{//}$ along $y$ direction ($B_y$).
\end{thebibliography}

\end{document}